\documentclass[english,reprint, aip, jcp, onecolumn]{revtex4-1}
\usepackage[T1]{fontenc}
\usepackage[latin9]{inputenc}
\usepackage{geometry}
\usepackage{babel}
\usepackage{textcomp}
\usepackage{amsmath}
\usepackage{stmaryrd}
\usepackage{graphicx}
\usepackage[unicode=true,
 bookmarks=true,bookmarksnumbered=false,bookmarksopen=false,
 breaklinks=true,pdfborder={0 0 0},pdfborderstyle={},backref=false,colorlinks=false]
 {hyperref}
\hypersetup{pdftitle={Efficient molecular density functional theory using generalized spherical harmonics expansions},
 pdfauthor={Lu Ding, Maximilien Levesque, Daniel Borgis and Luc Belloni}}

\makeatletter

\providecommand{\tabularnewline}{\\}
\newcommand{\lyxdot}{.}

\usepackage[T1]{fontenc}

\makeatother

\begin{document}

\title{Efficient molecular density functional theory using generalized spherical
harmonics expansions}

\author{Lu Ding}

\affiliation{Maison de la Simulation, USR 3441 CNRS-CEA-Universit� Paris-Saclay,
91191 Gif-sur-Yvette, France}

\author{Maximilien Levesque}

\affiliation{PASTEUR, D�partement de chimie, �cole normale sup�rieure, UPMC Univ. Paris
06, CNRS, PSL Research University, 75005 Paris, France}

\affiliation{Sorbonne Universit�s, UPMC Univ. Paris 06, �cole normale sup�rieure,
CNRS, Processus d'activation s�lective par transfert d'�nergie uni-�lectronique
ou radiatif (PASTEUR), 75005 Paris, France}

\author{Daniel Borgis}

\affiliation{Maison de la Simulation, USR 3441 CNRS-CEA-Universit� Paris-Saclay,
91191 Gif-sur-Yvette, France}

\affiliation{PASTEUR, D�partement de chimie, �cole normale sup�rieure, UPMC Univ. Paris
06, CNRS, PSL Research University, 75005 Paris, France}

\affiliation{Sorbonne Universit�s, UPMC Univ. Paris 06, �cole normale sup�rieure,
CNRS, Processus d'activation s�lective par transfert d'�nergie uni-�lectronique
ou radiatif (PASTEUR), 75005 Paris, France}

\author{Luc Belloni}

\affiliation{LIONS, NIMBE, CEA, CNRS, Universit� Paris-Saclay, 91191 Gif-sur-Yvette,
France}
\email{luc.belloni@cea.fr}

\date{\today}
\begin{abstract}
We show that generalized spherical harmonics are well suited for representing
the space and orientation molecular density in the resolution of the
molecular density functional theory. We consider the common system
made of a rigid solute of arbitrary complexity immersed in a molecular
solvent, both represented by molecules with interacting atomic sites
and classical force fields. The molecular solvent density $\rho(\mathbf{r},\mathbf{\Omega})$
around the solute is a function of the position $\mathbf{r}\equiv(x,y,z)$
and of the three Euler angles $\mathbf{\Omega}\equiv(\theta,\phi,\psi)$
describing the solvent orientation. The standard density functional,
equivalent to the HNC closure for the solute-solvent correlations
in the liquid theory, is minimized with respect to $\rho(\mathbf{r},\mathbf{\Omega})$.
The up-to-now very expensive angular convolution products are advantageously
replaced by simple products between projections onto generalized spherical
harmonics. The dramatic gain in speed of resolution enables to explore
in a systematic way molecular solutes of up to nanometric sizes in
arbitrary solvents and to calculate their solvation free energy and
associated microscopic solvent structure in at most a few minutes.
We finally illustrate the formalism by tackling the solvation of molecules
of various complexity in water.
\end{abstract}
\maketitle

\section{Introduction}

The knowledge of the free energy of solvation or chemical potential
of a molecular or macromolecular solute immersed in a molecular solvent
like water is the starting point of many applications in different
fields. Without surprise, beside experimental work, various numerical
theories/simulations have been developed following different directions
in order to predict such solvation free energy while minimizing the
restitution time. The atomic/molecular level of description where
the particles are described by sites interacting via classical force
fields (essentially Lennard-Jones and coulombic contributions) offers
a good compromise between expensive \emph{ab-initio }treatments (with
electronic, quantum mechanics description) and crude continuous solvent
models. The numerical difficulty originates from the large number
of solvent molecules to take into account. How to solve this statistical
mechanical problem? The molecular dynamics or Monte Carlo simulations
which explicitly consider up to millions of solvent molecules in a
simulation cell seem to be methods of choice for an exact resolution
but, in practice, are limited by prohibitive times to solution and
associated large statistical uncertainties. Consequently, there is
a clear demand for alternative theoretical routes. As usual, since
the beginning of the liquid state theory field in the 1950-1960's,
the approach based on the Ornstein-Zernike (OZ) equation, the integral
equations (IE) or the classical density functional theory (DFT) formalism
offers a good candidate for such calculations. \cite{hansen_theory_2013}
The goal is to derive the density of the solvent as a function of
its position and its orientation in the vicinity of the solute. For
polar solvent like water, the electrostatic couplings and resulting
hydrogen-bonding correlations are highly anisotropic and the angular
description requires a high level of sophistication. We briefly mention
here the Reference Interaction Site Model (RISM) approach which ignores
this full molecular analysis and replaces it by site-site correlations,
only; \cite{Chandler-RISM,hirata-rossky81,pettitt_integral_1982,pettitt07,pettitt08}
the gain in simplicity and speed is obvious since the interacting
particles are spherical; the price to pay is to deal with phenomenological
site-site OZ equation, correlation functions without proper statistical
mechanical foundation, and ad-hoc closures. RISM is well-developed
in its three-dimensional version \cite{hirata_molecular_2003,kloss_treatment_2008,maruyama_massively_2014,sergiievskyi_multigrid_2011,sergiievskyi_modelling_2012,kast-IAS2015}
and has provided valuable insight to a number of physical-chemistry
problems \cite{imai_locating_2006,yoshida_molecular_2009,casanova_evaluation_2007,kaminski_modeling_2010,kloss_quantum_2008,kast-IAS2015},
including the prediction of solvation free energies \cite{palmer_accurate_2010,sergiievskyi_3drism_2012,truchon_cavity_2014,sergiievskyi_solvation_2015,palmer2016,luchko2016,tielker-et-al2016}.
A RISM-based density functional theory has also been developed for
similar applications \cite{liu_site_2013,Wu_hydration14,Wu_hydration15}.
To bypass the limitations of the RISM approximation, and some of its
pitfalls, we propose here to stay at the more ambitious and demanding,
but otherwise more fundamental, full molecular level of description.
When the solvent and solute particles keep a simple shape, say 3-sites
${\rm H_{2}O}$ molecules around a spherical ion, it is natural to
express the solvent density $\rho(\mathbf{r},\mathbf{\Omega}_{12})$
in terms of solute-solvent separation $r$ and five Euler angles orientation.
This radial description has been studied in great details both in
bulk solvent and in solutions. Powerful formalisms which make use
of expansions onto rotational invariants or generalized spherical
harmonics enabled to solve the molecular Ornstein-Zernike equation
(MOZ) and integral equations for various densities, temperatures,
and compositions \cite{blum_invariant_1972,blum_invariant_1972-1,fries_solution_1985,richardi_molecular_1999,lombardero99,puibasset_bridge_2012,belloni14}.
This approach breaks down when the molecular/macromolecular solute
particle takes a complicated shape with many interacting sites. In
such case, it is desirable to consider the solvent density $\rho(\mathbf{r},\mathbf{\Omega})$
as a function of its 3D absolute position $\mathbf{r}\equiv(x,y,z)$
around the fixed solute and of its absolute orientation with respect
to a laboratory frame, characterized by three Euler angles $\mathbf{\Omega}\equiv(\theta,\phi,\psi)$.
Such approach was developed recently in a DFT framework (named MDFT,
for molecular density functional theory, in reference to MOZ) \cite{ramirez_density_2002,gendre_classical_2009,zhao_molecular_2011,borgis_molecular_2012,sergiievskyi_fast_2014}.
In the current implementation, the formalism requires as input the
full angular-dependent direct correlation function of the homogeneous
solvent - a difficult problem in itself \cite{ramirez_direct_2005,zhao_accurate_2013}
especially to get it precisely at all wave-lengths \cite{puibasset_bridge_2012,belloni-to-come}.
The computation of the excess free energy requires a double integration
over orientations for each spatial grid point, which made it prohibitive
to tackle large molecular systems. This computational limitation has
been overcome in special cases, such as the point-charge models of
water, for which the free-energy functional can be further approximated
and expressed in terms of two simpler fields than the full orientational
density, namely the density and polarization density fields. \cite{levesque_scalar_2012,jeanmairet_molecular_2013,jeanmairet_molecular_2013-1,jeanmairet_molecular_2015,jeanmairet_molecular_2016}

The objective of the present work is to develop a formalism and numerical
algorithms so efficient they unlock the resolution of 3D-DFT or OZ+IE
theories in the general case. Section II recalls the 3D-DFT approach
while Section III develops the formalism based on angular projections
onto carefully chosen basis of spherical harmonics. A few examples
applications are shown in Section IV.

\section{3D Molecular DFT and Ornstein-Zernike approach}

The goal is to derive the local and orientational molecular solvent
density $\rho(\mathbf{r},\mathbf{\Omega})$ where the vector $\mathbf{r}\equiv(x,y,z)$
defines the position of the rigid solvent molecule and $\mathbf{\Omega}\equiv(\theta,\phi,\psi)$
represents its orientation with respect to a fixed laboratory frame.
The direction of the main axis of the molecule is characterized by
the colatitude $\theta$ and longitude $\phi$ while $\psi$ is the
angle of rotation around this axis. The choice of solvent's origin
and main axis should take advantage of the molecular symmetry group.
For instance, for the water molecule of point group C$_{2\textrm{v}}$,
the origin is chosen at the oxygen site while the $z$ axis is its
$C_{2}$ main symmetry axis and points from the oxygen to the mid-point
between the hydrogens. It will be shown below that chosing high symmetry
axes implies notable simplifications.

The starting point of the liquid-state density functional theory (DFT)
consists in writing a functional $F\left[\rho(\mathbf{r},\mathbf{\Omega})\right]$
of the molecular density to be minimized. It is defined as the difference
between the grand potential of the solvated solute and the grand potential
of the homogeneous solvent at density $\rho_{\textrm{bulk}}$. It
is thus by definition the solvation free energy of the solute. Without
approximation for the moment, it may be splitted into ideal, external
and excess contributions\cite{evans92,evans_density_2009}:

\begin{equation}
F=F_{{\rm ideal}}+F_{{\rm ext}}+F_{{\rm excess}}\label{eq:Fid+Fext+Fexc}
\end{equation}

The ideal term, coming from the entropy of mixing of the solvent molecules,
reads

\begin{equation}
F_{{\rm ideal}}=k_{{\rm B}}T\iiint\mathrm{d}\mathbf{r}\iiint\mathrm{d}\mathbf{\Omega}\left[\rho(\mathbf{r},\mathbf{\Omega})\ln\dfrac{\rho(\mathbf{r},\mathbf{\Omega})}{\rho_{{\rm bulk}}}-\Delta\rho(\mathbf{r},\mathbf{\Omega})\right]\label{eq:1.1}
\end{equation}
where $T$ is the temperature, $k_{\textrm{B}}$ is the Boltzmann
constant, $k_{{\rm B}}T$ is the thermal energy, and $\Delta\rho(\mathbf{r},\mathbf{\Omega})\equiv\rho(\mathbf{r},\mathbf{\Omega})-\rho_{{\rm bulk}}$
with $\rho_{\textrm{bulk}}\equiv n_{\textrm{bulk}}/8\pi\text{\texttwosuperior}$.
$n_{\textrm{bulk}}$ is the bulk density. The external contribution
comes from the interaction potential $V_{{\rm ext}}$ between the
solute molecule and one solvent molecule:
\begin{equation}
F_{{\rm ext}}=\iiint\mathrm{d}\mathbf{r}\iiint\mathrm{d}\mathbf{\Omega}\rho(\mathbf{r},\mathbf{\Omega})V_{{\rm ext}}(\mathbf{r},\mathbf{\Omega}).
\end{equation}
In the usual case of spherically symmetric site-site interaction potentials,
$V_{{\rm ext}}$ reads
\begin{equation}
V_{{\rm ext}}(\mathbf{r},\mathbf{\Omega})=\sum_{i={\rm solvent\,site}}\sum_{j={\rm solute\,site}}v_{ij}\left(\left|\mathbf{r}+\mathbf{s}_{i}(\mathbf{\Omega})-\mathbf{r}_{j}\right|\right)\label{eq:1.3}
\end{equation}
where $\mathbf{s}_{i}$ is the intra vector joining the solvent origin
to the site $i$. When the DFT is solved inside a cubic cell of edge
$L$ with periodic boundary conditions, the contributions from the
neighboring solute images must be added to Eq. \ref{eq:1.3} in the
obvious and usual way. The coulombic $1/r$ contribution to the external
potential is derived by solving the Poisson equation inside the cell.
Again, this imposed potential is constant in what follows.

The final, excess term involves the correlations between the solvent
molecules perturbed by the neighboring solute. As usual in such liquid-state
theory, an approximation must be assumed for this contribution. The
bare, well developed and documented functional, first term in an infinite
Taylor expansion around the liquid bulk density, reads:
\begin{equation}
\beta F_{{\rm excess}}=-\frac{1}{2}\iiint\mathrm{d}\mathbf{r}_{1}\iiint\mathrm{d}\mathbf{\Omega}_{1}\iiint\mathrm{d}\mathbf{r}_{2}\iiint\mathrm{d}\mathbf{\Omega}_{2}\Delta\rho(\mathbf{r}_{1},\mathbf{\Omega}_{1})c(\mathbf{r}_{12},\mathbf{\Omega}_{1},\mathbf{\Omega}_{2})\Delta\rho(\mathbf{r}_{2},\mathbf{\Omega}_{2})\label{eq:1.4}
\end{equation}
where $c(\mathbf{r}_{12},\mathbf{\Omega}_{1},\mathbf{\Omega}_{2})$
is the bulk solvent-solvent molecular direct correlation function
(DCF), which depends on the distance $r_{12}$ between the two solvent
molecules and the five Euler angles characterizing their relative
orientation (invariant by translation and rotation of the ensemble
$(\mathbf{r}_{12},\mathbf{\Omega}_{1},\mathbf{\Omega}_{2})$ with
respect to the fixed frame). We remind the reader that even if three
Euler angles are necessary to define the orientation of a single molecule,
five only are necessary for defining \emph{relative }orientations.
The function $c$ of the bulk solvent for a given temperature and
pressure is an input in the present approach and is provided by previous
extensive Monte Carlo + IE bulk calculations \cite{puibasset_bridge_2012,belloni14}.

The formal functional differentiation of \ref{eq:1.1} leads to:
\begin{equation}
\rho(\mathbf{r},\mathbf{\Omega})=\rho_{{\rm bulk}}\exp\left[-\beta V_{{\rm ext}}(\mathbf{r},\mathbf{\Omega})+\gamma(\mathbf{r},\mathbf{\Omega})\right]\label{eq:1.5}
\end{equation}
where $\beta=1/k_{{\rm B}}T$ and $\gamma(\mathbf{r},\mathbf{\Omega})$
represents the indirect (total minus direct) solute-solvent correlation
function which is related to the previous functions via the solute-solvent
Ornstein-Zernike equation:
\begin{equation}
\gamma(\mathbf{r}_{1},\mathbf{\Omega}_{1})=\iiint\mathrm{d}\mathbf{r}_{2}\iiint\mathrm{d}\mathbf{\Omega}_{2}c(\mathbf{r}_{12},\mathbf{\Omega}_{1},\mathbf{\Omega}_{2})\Delta\rho(\mathbf{r}_{2},\mathbf{\Omega}_{2})\label{eq:1.6}
\end{equation}

The integral equation \ref{eq:1.5} is nothing but the HNC approximation
for the solute-solvent correlations, which ignores the so-called bridge
function. Inclusion of more sophisticated excess functionals or bridge
functions will be investigated in future works. 

The excess free energy functional \ref{eq:1.4} may be written as:
\begin{equation}
F_{{\rm excess}}=-\frac{1}{2}\iiint\mathrm{d}\mathbf{r}_{1}\iiint\mathrm{d}\mathbf{\Omega}_{1}\Delta\rho(\mathbf{r}_{1},\mathbf{\Omega}_{1})\gamma(\mathbf{r}_{1},\mathbf{\Omega}_{1}).
\end{equation}
In practice, the numerical resolution consists in general to describe
the cubic cell with a 3D grid of $N\times N\times N$ spatial positions
(grid nodes) and mesh size $L/N$. $N$ is typically below 256 for
computer memory reasons. Generalization to parallelepiped cells or
different directional mesh resolutions is straightforward. For each
of the $N^{3}$ grid points, the orientation is characterized by different
$\Omega$ triplets. For simplicity, we use $N_{\theta}$, $N_{\phi}$,
$N_{\psi}$ decoupled values $\theta_{i}$, $\phi_{j}$, $\psi_{k}$
chosen from the Gauss quadrature. In general, $0\le\theta_{i}<\pi$,
$0\le\phi_{j}<2\pi$ and $0\le\psi_{k}<2\pi$. In the case of the
${\rm H_{2}O}$ molecule (of symmetry group $\mathrm{C}_{2v}$), $0\le\psi_{k}<\pi$
is sufficient. Typical numbers are $5-10$ for each of the three angles.

The resolution consists either to numerically minimize the total DFT
functional \ref{eq:1.1} with respect to the solvent density $\text{\ensuremath{\Delta\rho}(\ensuremath{\mathbf{r}},\ensuremath{\mathbf{\Omega}})}$
or, equivalently, to solve the integral equation \ref{eq:1.5}. We
choose the former route in the present study. The process is iterative.
At convergence, $\text{\ensuremath{\Delta\rho}(\ensuremath{\mathbf{r}},\ensuremath{\mathbf{\Omega}})}$
gives the equilibrium solvent profiles around the solute and the value
taken by $F$ provides the free energy of solvation.

The most demanding and challenging part of the calculation is obviously
the excess part \ref{eq:1.4} or \ref{eq:1.6} which requires a 6D
spatial+angular convolution. The spatial one is naturally performed
in the Fourier space where \ref{eq:1.6} becomes: 
\begin{equation}
\hat{\gamma}(\mathbf{q},\mathbf{\Omega}_{1})=\iiint\mathrm{d}\mathbf{\Omega}_{2}\hat{c}(\mathbf{q},\mathbf{\Omega}_{1},\mathbf{\Omega}_{2})\Delta\hat{\rho}(\mathbf{q},\mathbf{\Omega}_{2}).\label{eq:1.8}
\end{equation}
$\mathbf{q}\equiv(q_{x},q_{y},q_{z})$ is the vector in the Fourier
space, each component $q_{i}$ is discretized in $N$ values multiples
of $2\pi/L$. The hat functions indicate the 3D Fourier transformed
functions, defined as $\hat{f}(\mathbf{q},\mathbf{\Omega})=\iiint f(\mathbf{r},\mathbf{\Omega})e^{i\mathbf{q}\cdot\mathbf{r}}\mathrm{d}\mathbf{r}$.
They are complex quantities. Of course, we use state-of-the-art FFT
libraries to compute the space convolution with a complexity in $\mathcal{O}\left(N\log N\right)$
instead of the $\mathcal{O}\left(N^{2}\right)$ in the naive implementation.The
angular convolution which remains in the MOZ equation \ref{eq:1.8},
when implemented straightforwardly in refs \cite{gendre_classical_2009,zhao_molecular_2011},
represents the main barrier for an efficient resolution: for each
of the $N^{3}$ values of $\boldsymbol{q}$ and for each of the orientation
triplet $\mathbf{\Omega}_{1}$, one must perform a 3D integral over
the whole orientation triplet $\mathbf{\Omega}_{2}$ using angular
quadratures! Indeed, even the naive implementation is not so straightforward
in practice since, expressed in the laboratory frame, the 8-variables
angular DCF $\hat{c}(\mathbf{q},\boldsymbol{\Omega}_{1},\boldsymbol{\Omega}_{2})$
is too large to be stored. This problem can be solved by storing the
DCF in the so-called intermolecular frame, for which the $z$-axis
is taken in the direction of $\mathbf{q}$, so that $\hat{c}(q,\boldsymbol{\Omega}_{1}^{'},\boldsymbol{\Omega}_{2}^{'})$
can be expressed as a function of only 6 variables when accounting
from rotational invariance around $\mathbf{q}$. For each value of
$\mathbf{q}$, one thus needs also to infer the correspondence between
orientations $\boldsymbol{\Omega}_{i}$ and $\boldsymbol{\Omega}_{i}^{'}(\mathbf{q},\boldsymbol{\Omega}_{i})$
in the fixed and molecular frame, respectively. This process, whatever
the algorithm (storing or recomputing), further impairs the numerical
efficiency.

In the next Section, we show that the use of expansions onto basis
of generalized spherical harmonics will (i) advantageously replace
the angular convolution by simple products between projections, and
(ii) reduce the memory footprint of the storage of the DCF.

\section{Expansion onto generalized spherical harmonics}

The angular dependency of the solvent density $\Delta\rho(\mathbf{r},\mathbf{\Omega})$
is expanded for each point $\mathbf{r}$ of the 3D network onto a
basis of carefully chosen functions, the generalized spherical harmonics
$R_{\mu'\mu}^{m}(\mathbf{\Omega})$ following Messiah and Blum's notations
\cite{messiah_tome2,blum_invariant_1972}:
\begin{equation}
\Delta\rho(\mathbf{r},\mathbf{\Omega})=\sum_{m=0}^{n_{\max}}\sum_{\mu'=-m}^{m}\sum_{\mu=-m}^{m}f_{m}\Delta\rho_{\mu'\mu}^{m}(\mathbf{r})R_{\mu'\mu}^{m}(\mathbf{\Omega}),\label{eq:1.9}
\end{equation}
with
\begin{equation}
R_{\mu'\mu}^{m}(\mathbf{\Omega})=r_{\mu'\mu}^{m}(\theta)e^{-i\mu'\phi-i\mu\psi},
\end{equation}
where $r_{\mu'\mu}^{m}(\theta)$ is the generalized Legendre polynomial
and $f_{m}=\sqrt{2m+1}$ is a normalization factor. Each labelled
coefficient, the so-called projections, in the sum \ref{eq:1.9} is
obtained by angular integral of the original function (projection
onto the corresponding basis vector):
\begin{equation}
\Delta\rho_{\mu'\mu}^{m}(\mathbf{r})=f_{m}\iiint\Delta\rho(\mathbf{r},\mathbf{\Omega})R_{\mu'\mu}^{m*}(\mathbf{\Omega})\mathrm{d}\mathbf{\Omega}.\label{eq:1.11}
\end{equation}
The expansion in \ref{eq:1.9} is in principle infinite. In practice,
it is truncated at $m\leq n_{\max}$, which defines the basis$\left\{ n_{\max}\right\} $
of angular functions. In order to be consistent with the prescription
of the Gauss quadrature, the number of angles for $\theta$, $\phi$,
and $\psi$ will verify $N_{\theta}=n_{\max}+1$, $N_{\phi}=2n_{\max}+1$
and $N_{\psi}=2\left(n_{\textrm{max}}/s\right)+1$ where $s$ is the
order of the symmetry axis used as main molecular axis for the solvent
molecule ($s=2$ for $C_{2V}$ molecules like water) and the division
is an integer division. Since the input function $\Delta\rho(\mathbf{r},\mathbf{\Omega})$
is \textit{real}-\emph{valued}, a symmetry relation follows between
the complex-valued projections $\Delta\rho_{\mu'\mu}^{m}(\mathbf{r})$:
\begin{equation}
\Delta\rho_{\underline{\mu'}\underline{\mu}}^{m}(\mathbf{r})=\left(-1\right)^{\mu'+\mu}\Delta\rho_{\mu'\mu}^{m*}(\mathbf{r}),\label{eq:1.12}
\end{equation}
where $\underbar{\ensuremath{\mu}}\equiv-\mu$. As a consequence,
it is sufficient to deal here with $\mu'\geq0$ (or $\mu\geq0$).
For ${\rm H_{2}O}$ solvent, $\mu$ is even and the total number of
independent projections per spacial grid node is 4, 19, 40, 85, 140
for $n_{\max}=1$, 2, 3, 4 and 5, as shown in table \ref{tab:table_norientations_nprojections}.

The transformation from $\text{\ensuremath{\Delta\rho}(\ensuremath{\mathbf{r}},\ensuremath{\mathbf{\Omega}})}$
to $\Delta\rho_{\mu'\mu}^{m}(\mathbf{r})$ through Eq. \ref{eq:1.9}
and \ref{eq:1.11} is numerically performed using a fast 3-step algorithm
\cite{lado_95} described in the appendix. Each $r$-projection is
then Fourier transformed by FFT
\begin{equation}
\Delta\hat{\rho}_{\mu'\mu}^{m}(\mathbf{q})=\iiint\Delta\rho_{\mu'\mu}^{m}(\mathbf{r})e^{i\mathbf{q}\cdot\mathbf{r}}\mathrm{d}\mathbf{r}.
\end{equation}
Of course, since the angle $\Omega$ is defined with respect to a
fixed frame, independent of $\mathbf{r}$, this means that 
\begin{equation}
\Delta\hat{\rho}(\mathbf{q},\mathbf{\Omega})=\sum_{m=0}^{n_{\max}}\sum_{\mu'=-m}^{m}\sum_{\mu=-m}^{m}f_{m}\Delta\hat{\rho}_{\mu'\mu}^{m}(\mathbf{q})R_{\mu'\mu}^{m}(\mathbf{\Omega}),\label{eq:1.13-1}
\end{equation}
and the symmetry relation \ref{eq:1.12} becomes:
\begin{equation}
\Delta\hat{\rho}_{\underline{\mu'}\underline{\mu}}^{m}(\mathbf{q})=\left(-1\right)^{\mu'+\mu}\Delta\hat{\rho}_{\mu'\mu}^{m*}(-\mathbf{q}),\label{eq:1.14}
\end{equation}
which halves the number of $\mathbf{q}$ values to consider.

In the same way, the bulk function can be decomposed into:
\begin{equation}
\hat{c}(\mathbf{q},\mathbf{\Omega}_{1},\mathbf{\Omega}_{2})=\sum_{mnl\mu\nu}\hat{c}_{\mu\nu}^{mnl}(q)\Phi_{\mu\nu}^{mnl}(\hat{\mathbf{q}},\mathbf{\Omega}_{1},\mathbf{\Omega}_{2}),
\end{equation}
where the coefficients $c_{\mu\nu}^{mnl}\left(q\right)$ depend here
on the norm $q$ only and the rotational invariants are defined such
as to verify the invariance by rotation of the ensemble:
\begin{equation}
\Phi_{\mu\nu}^{mnl}(\hat{\mathbf{q}},\mathbf{\Omega}_{1},\mathbf{\Omega}_{2})=f_{m}f_{n}\sum_{\mu'\nu'\lambda'}\left(\begin{array}{ccc}
m & n & l\\
\mu' & \nu' & \lambda'
\end{array}\right)R_{\mu'\mu}^{m}(\mathbf{\Omega}_{1})R_{\nu'\nu}^{n}(\mathbf{\Omega}_{2})R_{\lambda'0}^{l}(\hat{\mathbf{q}}).\label{eq:1.16}
\end{equation}
The coefficients $\left(\begin{array}{ccc}
m & n & l\\
\mu' & \nu' & \lambda'
\end{array}\right)$ are the usual 3-j-symbols. The complex projections $\hat{c}_{\mu\nu}^{mnl}$
verify symmetry relations because $c$ is a real-valued function and
the solvent molecules 1, 2 are identical:
\begin{equation}
\hat{c}_{\underline{\mu}\underline{\nu}}^{mnl}=(-1)^{m+n+\mu+\nu}\hat{c}_{\mu\nu}^{mnl*}
\end{equation}
\begin{equation}
\hat{c}_{\nu\mu}^{nml}=(-1)^{m+n}\hat{c}_{\mu\nu}^{mnl}
\end{equation}
In the case of ${\rm H_{2}O}$ symmetry, $\mu$ and $\nu$ are even
and $\hat{c}_{\mu\nu}^{mnl}$ is real-valued if $l$ is even and pure
imaginary if $l$ is odd. Consequently:
\begin{equation}
\hat{c}_{\underline{\mu}\underline{\nu}}^{mnl}=(-1)^{m+n}\hat{c}_{\mu\nu}^{mnl*}=(-1)^{m+n+l}\hat{c}_{\mu\nu}^{mnl}.
\end{equation}
In that case, the number of independent real coefficients is 4, 27,
79, 250, 549 for $n_{\max}=1,2,3,4,5$, respectively.

What is the interest of all these projections? The angular integral
over $\boldsymbol{\Omega}_{2}$ in equation \ref{eq:1.8} now concerns
only two spherical harmonics $R(\boldsymbol{\Omega}_{2})$ in equations
\ref{eq:1.13-1} and \ref{eq:1.16}: it can now be performed analytically! 

The calculation is again simplified and accelerated by switching to
the local, molecular frame linked to $\hat{q}$, taken as principal
axis. The orientation of the solvent molecule in this frame is noted
$\mathbf{\Omega'}$. Composition relations between spherical harmonics
during the transformation (rotation) from fixed to local frames are
simple matrix products, one for each $m$ indices:
\begin{equation}
\mathbf{R}^{m}(\mathbf{\Omega})=\mathbf{R}^{m}(\hat{\mathbf{q}})\mathbf{R}^{m}(\mathbf{\Omega'})\label{eq:22}
\end{equation}
\begin{equation}
R_{\mu'\mu}^{m}(\mathbf{\Omega})=\sum_{\chi}R_{\mu'\chi}^{m}(\hat{\mathbf{q}})R_{\chi\mu}^{m}(\mathbf{\Omega'})
\end{equation}
In the local frame, the expansion analogous to equation \ref{eq:1.13-1}
becomes:
\begin{equation}
\Delta\hat{\rho}(\mathbf{q},\mathbf{\Omega'})=\sum_{m\mu\chi}f_{m}\Delta\hat{\rho}_{\mu;\chi}^{m}(\mathbf{q})R_{\chi\mu}^{m}(\mathbf{\Omega'}).\label{eq:1.20}
\end{equation}
 The new coefficients (be careful of the new lower indices notation
consistent with Blum's) are deduced from the previous ones by a transformation
analogous of the so-called $\chi$-transform of Blum\cite{blum_invariant_1972-1,blum_invariant_1972}:
\begin{equation}
\Delta\hat{\rho}_{\mu;\chi}^{m}(\mathbf{q})=\sum_{\mu'}\Delta\hat{\rho}_{\mu'\mu}^{m}(\mathbf{q})R_{\mu'\chi}^{m}(\hat{\mathbf{q}}).\label{eq:1.21}
\end{equation}
For each discrete value of $\mathbf{q}$, the ensemble of $R_{\mu'\chi}^{m}(\hat{\mathbf{q}})$
projections is calculated using fast recurrence relations depending
only on the Cartesian coordinates of $\mathbf{q}$\cite{choi_rotmat_1999}.
Moreover, $\mathbf{q}$ and $-\mathbf{q}$ require a single treatment
since
\begin{equation}
R_{\mu'\chi}^{m}(-\hat{\mathbf{q}})=\left(-1\right)^{m}R_{\mu'\underline{\chi}}^{m}(\hat{\mathbf{q}})=\left(-1\right)^{m+\mu'+\chi}R_{\underline{\mu'}\chi}^{m}(\hat{\mathbf{q}})
\end{equation}
In notation $\chi$, the general symmetry relation of equation \ref{eq:1.14}
becomes:
\begin{equation}
\Delta\hat{\rho}_{\underline{\mu'};\chi}^{m}(\mathbf{q})=\left(-1\right)^{m+\mu'+\chi}\Delta\hat{\rho}_{\mu';\chi}^{m*}(-\mathbf{q})
\end{equation}
In the same way, the bulk $\hat{c}$ function reads in this new frame\cite{blum_invariant_1972}:
\begin{equation}
\hat{c}(q,\mathbf{\Omega'}_{1},\mathbf{\Omega'}_{2})=\sum_{mn\mu\nu\chi}f_{m}f_{n}\hat{c}_{\mu\nu;\chi}^{mn}(q)R_{\chi\mu}^{m}(\mathbf{\Omega'}_{1})R_{\underline{\chi}\nu}^{n}(\mathbf{\Omega'}_{2})\label{eq:1.24}
\end{equation}
where the new coefficients are deduced from the old ones through the
Blum's \textquotedbl{}$\chi$-transform\textquotedbl{}: 
\begin{equation}
\hat{c}_{\mu\nu;\chi}^{mn}(q)=\sum_{\chi}\left(\begin{array}{ccc}
m & n & l\\
\chi & \underline{\chi} & 0
\end{array}\right)\hat{c}_{\mu\nu}^{mnl}(q)
\end{equation}
Some symmetry relations apply, even for molecules without symmetry:

\begin{equation}
\hat{c}_{\underline{\mu}\underline{\nu};\chi}^{mn}=\left(-1\right)^{m+n+\mu+\nu}\hat{c}_{\mu\nu;\chi}^{mn*},
\end{equation}
and
\begin{equation}
\hat{c}_{\nu\mu;\chi}^{nm}=\left(-1\right)^{m+n}\hat{c}_{\mu\nu;\chi}^{mn}.
\end{equation}
In the specific case of water,
\begin{equation}
\hat{c}_{\mu\nu;\underline{\chi}}^{mn}=\hat{c}_{\underline{\mu}\underline{\nu};\chi}^{mn}=\left(-1\right)^{m+n}\hat{c}_{\mu\nu;\chi}^{mn*}.
\end{equation}

Finally, the insertion of expansions \ref{eq:1.20} and \ref{eq:1.24}
into the OZ convolution product \ref{eq:1.8} (formally valid for
any reference frame, so in particular for the local one) followed
by an analytical integration over $\mathbf{\Omega'}_{2}$ (thanks
to the orthogonality of the spherical harmonics) leads to a very simple
OZ relation between $\hat{c}$, $\Delta\hat{\rho}$ and $\hat{\gamma}$
$\chi$-projections:
\begin{equation}
\hat{\gamma}_{\mu;\chi}^{m}(\mathbf{q})=\sum_{n\nu}\left(-1\right)^{\chi+\nu}\hat{c}_{\mu\nu;\chi}^{mn}(q)\Delta\hat{\rho}_{\underline{\nu};\chi}^{n}(\mathbf{q})\label{eq:1.28}
\end{equation}
This OZ relation constitutes the main result of the present formalism
and manuscript. It replaces the expensive angular convolution product
\ref{eq:1.8} by simple algebraic products between projections in
the local frame! This can be seen as the angular analogous of the
replacement of spatial convolution in equation \ref{eq:1.6} by direct
product in Fourier space \ref{eq:1.8}. It is important to note that
different $\chi$ values do not mix in \ref{eq:1.28}; there is one
simple matrix multiplication for each $\chi$ value. 

Once the $\hat{\gamma}_{\chi}$ projections have been derived from
the OZ equation, the return to the laboratory frame follows a relation
inverse of equation \ref{eq:1.20}:
\begin{equation}
\hat{\gamma}_{\mu'\mu}^{m}(\mathbf{q})=\sum_{\chi}\hat{\gamma}_{\mu;\chi}^{m}(\mathbf{q})R_{\mu'\chi}^{m*}(\hat{\mathbf{q}})\label{eq:1.29}
\end{equation}
Note that the transformation between fixed and local frames in \ref{eq:1.21}
and \ref{eq:1.29} invoke the spherical harmonics $R(\hat{\mathbf{q}})$,
where $(\hat{\mathbf{q}})$ is understood as the rotation which goes
from the fixed to the local frames. This last one is not defined univocally
because there is freedom in the choice of the rotation angle around
$\hat{\mathbf{q}}$. Fortunately, it is satisfying to verify in the
previous analysis that this angle is completely irrelevant in the
final result: indeed, the ensemble \ref{eq:1.21}, \ref{eq:1.28},
\ref{eq:1.29} involves products of the form $R_{?\chi}^{?}(\hat{\mathbf{q}})R_{?\chi}^{?*}(\hat{\mathbf{q}})$
which are really independent of it.

Finally, we apply an inverse FFT to $\hat{\gamma}_{\mu'\mu}^{m}(\mathbf{q})$,
then gather all projections of $\gamma_{\mu'\mu}^{m}(\mathbf{r})$
like in equation \ref{eq:1.9}. We end up with the desired indirect
correlation function $\gamma(\mathbf{r},\mathbf{\Omega})$:
\begin{equation}
\gamma(\mathbf{r},\mathbf{\Omega})=\sum_{m\mu'\mu}f_{m}\gamma_{\mu'\mu}^{m}(\mathbf{r})R_{\mu'\mu}^{m}(\mathbf{\Omega})\label{eq:1.30}
\end{equation}
The very expensive original OZ equation \ref{eq:1.6} has thus been
replaced by the series of cheap steps \ref{eq:1.11}, \ref{eq:1.13-1},
\ref{eq:1.21}, \ref{eq:1.28}, \ref{eq:1.29}, \ref{eq:1.30}:
\begin{equation}
\Delta\rho(\mathbf{r},\mathbf{\Omega})\rightarrow\Delta\rho_{\mu'\mu}^{m}(\mathbf{r})\rightarrow\Delta\hat{\rho}_{\mu'\mu}^{m}(\mathbf{q})\rightarrow\Delta\hat{\rho}_{\mu;\chi}^{m}(\mathbf{q})\rightarrow\hat{\gamma}_{\mu;\chi}^{m}(\mathbf{q})\rightarrow\hat{\gamma}_{\mu'\mu}^{m}(\mathbf{q})\rightarrow\gamma_{\mu'\mu}^{m}(\mathbf{r})\rightarrow\gamma(\mathbf{r},\mathbf{\Omega}).
\end{equation}

\section{Implementation and examples}

We apply the present DFT approach in the case of the SPC/E model of
water. The ${\rm H_{2}O}$ solvent molecule is characterized by one
LJ site localized at the O site and three partial charges at the O,
H, H sites. For this model, the DCF projections $c_{\mu\nu;\chi}^{mn}(q)$
at different orders of accuracy $n_{max}$ have been previously obtained
by combining Monte Carlo simulation data at short distances and HNC
closure at long distances and solving the resulting 1D MOZ+mixed IE.
The temperature is 298.15 K and the bulk density is 997 g/L \cite{puibasset_bridge_2012,belloni-to-come}.
. 

As mentioned above, the functional of equation \ref{eq:Fid+Fext+Fexc}
is minimized with respect to $\rho(\mathbf{r},\mathbf{\boldsymbol{\Omega}})$
using the quasi-Newton minimizer L-BFGS\cite{BFGS}. The density is
usually initiated at $\rho_{\textrm{bulk}}\exp\left(-\beta V_{ext}\left(\mathbf{r},\mathbf{\boldsymbol{\Omega}}\right)\right)$.
LBFGS requires at each minimization step the value of the functional
and its gradient
\begin{equation}
\frac{\beta\delta F}{\delta\rho(\mathbf{r},\boldsymbol{\Omega})}=\log\left(\frac{\rho(\mathbf{r,}\boldsymbol{\Omega})}{\rho_{bulk}}\right)+\beta V_{ext}(\mathbf{r},\boldsymbol{\Omega})-\gamma(\mathbf{r},\boldsymbol{\Omega}).\label{eq:gradient}
\end{equation}

\begin{figure}
\begin{centering}
\includegraphics[width=8.5cm]{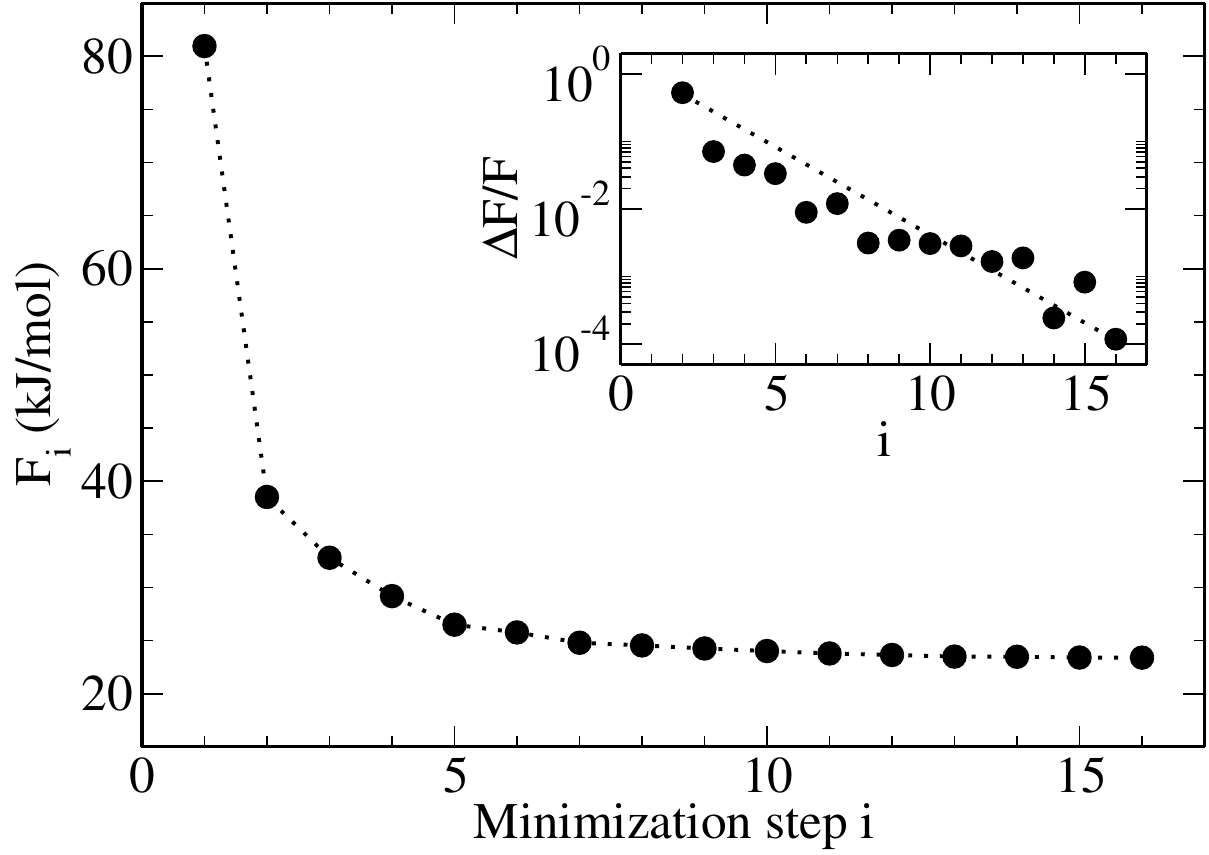}
\par\end{centering}
\caption{Convergence of the free energy estimation during the minimization
process for a CO$_{2}$ molecule in water with a box size $L=24$~\AA,
$N=72$ and $n_{max}=3$. The inset shows the evolution of the relative
difference in the free energy functional between successive steps.
\label{fig:convergence}}
\end{figure}

A typical minimization process using the new method described above
for computing $\gamma(\mathbf{r},\boldsymbol{\Omega})$ is illustrated
in Fig.~\ref{fig:convergence} for the CO$_{2}$ molecule in water;
it is seen that in this case a relative error of $10^{-4}$ is reached
after only $\approx20$ cycles. To our experience, convergence is
reached before $\approx35$ steps or never.

We begin by comparing the numerical efficiency of the new method to
that of the original, direct method, described by equation \ref{eq:1.6}. 

The old, direct method requires to pre-compute and store the DCF in
local frame, $\hat{c}(q,\mathbf{\boldsymbol{\Omega}'}_{1},\mathbf{\boldsymbol{\Omega}'}_{2})$
using eq.~\ref{eq:1.24} and then involves three successive steps,
namely (i) to fast Fourier transform $\Delta\rho(\mathbf{r,\mathbf{\boldsymbol{\Omega})}}$
to $\Delta\rho(\mathbf{q,\mathbf{\boldsymbol{\Omega})}}$, (ii) to
compute $\gamma(\mathbf{q,\mathbf{\boldsymbol{\Omega})}}$ using the
MOZ equation~\ref{eq:1.8} in $q$ space with the stored DCF, and
finally (iii) to inverse fourier Transform to $\gamma(\mathbf{r,\mathbf{\boldsymbol{\Omega})}}$
. 

On the other side, the new method involves the succession of seven
steps described in the previous section. Note that steps 1-2 and 6-7,
i.e. angular transforms and spatial transforms, could well be inverted.
However it is more efficient to go with the angular transforms first,
since for a given $n_{\textrm{max}}$ there are less projections than
angles (see table~\ref{tab:table_norientations_nprojections}), and
thus less functions to Fourier transform.

\begin{table}
\begin{centering}
\begin{tabular}{|c|c|c|c|c|c|c|}
\hline 
 & \multicolumn{3}{c|}{for generic solvent molecules} & \multicolumn{3}{c}{for $C_{2V}$ molecules like H$_{2}$O}\tabularnewline
\hline 
$n_{\textrm{max}}$ & $N_{\boldsymbol{\Omega}}$ & $N_{\text{complex-valued projections}}$ & $N_{\text{independent real-valued projections}}$ & $N_{\boldsymbol{\Omega}}$ & $N_{\text{complex-valued projections}}$ & $N_{\text{independent real-valued projections}}$\tabularnewline
\hline 
\hline 
1 & 18 & 10 & 7 & 6 & 4 & 4\tabularnewline
\hline 
2 & 75 & 35 & 22 & 45 & 19 & 14\tabularnewline
\hline 
3 & 196 & 84 & 50 & 84 & 40 & 28\tabularnewline
\hline 
4 & 405 & 165 & 95 & 225 & 85 & 55\tabularnewline
\hline 
5 & 726 & 286 & 161 & 330 & 140 & 88\tabularnewline
\hline 
\end{tabular}
\par\end{centering}
\caption{Correspondance between the number of orientations, $N_{\Omega}=N_{\theta}\times N_{\phi}\times N_{\psi}$,
and the number of projections for generic solvent molecules and for
water. The number of independent real projections uses the symmetry
rule from equation \ref{eq:1.12}. For C$_{2V}$ molecules, we add
the constraint that $\psi$ lies in $[0,\pi[$ and not $[0,2\pi[$
and that $\mu$ is even.\label{tab:table_norientations_nprojections}}
\end{table}

\begin{figure}
\begin{centering}
\includegraphics[width=8.5cm]{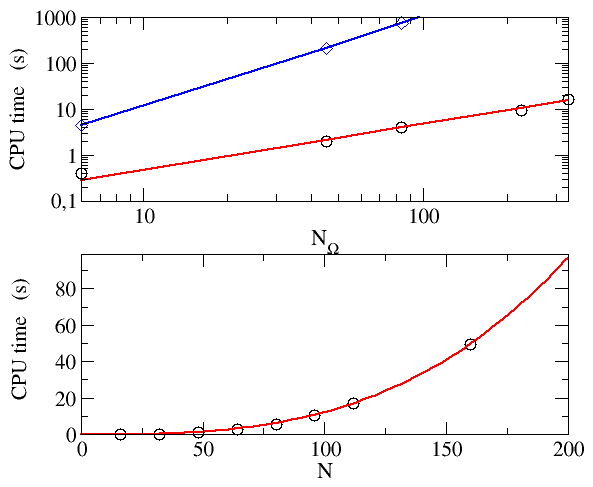}
\par\end{centering}
\caption{(Top) CPU time for the computation of $F_{exc}$ within a single minimization
step using the direct algorithm (blue diamonds) and the new one (black
circles) with $N=72$ as a function of number of discrete orientations
per spatial grid point, $N_{\Omega}$. The displayed numbers $N_{\Omega}=6$,
45, 84, 225, 330 correspond to $n_{max}=1$, 2, 3 ,4, 5, respectively.
The red and blue lines represent the best fit to linear behavior,
$T=aN_{\Omega}$, and quadratic behavior, $T=bN_{\Omega}^{2}$ , respectively.
(Bottom) same quantity as function of spatial grid point number $N$
using the new algorithm with $n_{max}=3$. The red line represents
the best fit to cubic behavior $T=cN^{3}$.\label{fig:CPU_vs_N}}
 
\end{figure}

We show in Fig.~\ref{fig:CPU_vs_N} that the CPU time requested to
compute $F_{exc}$ and $\gamma(\mathbf{q,\mathbf{\boldsymbol{\Omega})}}$
is indeed much lower with the new algorithm and that it behaves linearly
with respect to the chosen number of orientations per grid point,
$N_{\Omega}$, whereas it is quadratic in the direct algorithm. The
numerical gain is of a factor 200 for $n_{max}=3$ (84 orientations)
and 750 for $n_{max}=5$ (330 orientations). No surprise, the dependence
with respect to the number of spatial grid points $N$ is clearly
cubic, as shown in the bottom panel of Fig.\ref{fig:CPU_vs_N}. The
quoted CPU times refer to calculations on a single thread on an INTEL
Sandy Bridge $\varcopyright$ processor at 2 GHz. No parallelism of
whatever kind is included here. The important information to take
from that last figure is that even in that single-thread case, and
for $n_{max}=3$ , the calculation of $\gamma(\mathbf{q,\mathbf{\boldsymbol{\Omega})}}$
takes a few seconds for a grid of size $72^{3}$ , and above a minute
for $200^{3}$.

\begin{figure}
\begin{centering}
\includegraphics[width=8.5cm]{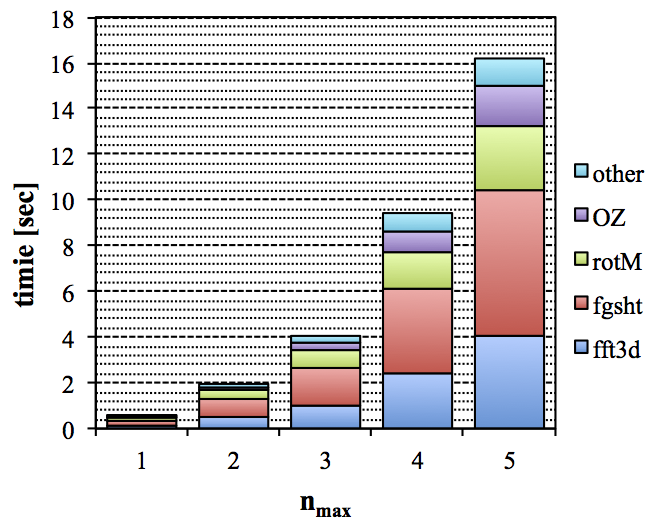}
\par\end{centering}
\caption{Decomposition of the CPU time of the different steps involved in the
calculation of $\gamma(\mathbf{r},\boldsymbol{\Omega})$ from $\Delta\rho(\mathbf{r},\boldsymbol{\Omega})$
at different angular resolutions for different $n_{\textrm{max}}$:
Fast generalized spherical harmonics transforms (red, steps $1+7$);
3D-FFT (blue, steps $2+6$); rotations between laboratory and local
frames (green, steps $3+5$); and resolution of the molecular Ornstein-Zernike
equation in the local frame (purple, step $4$).\label{fig:CPU-decomposition}}
\end{figure}

In Fig.~\ref{fig:CPU-decomposition} we show the decompostion of
the CPU time along the different steps of the algorithm for different
$n_{\textrm{max}}$: Although the Fast Generalized Spherical Harmonics
Transform (FGSHT) is the most time-consuming, the different steps
are rather equilibrated. None of them is a bottleneck.

\begin{figure}
\begin{centering}
\includegraphics[width=8.5cm]{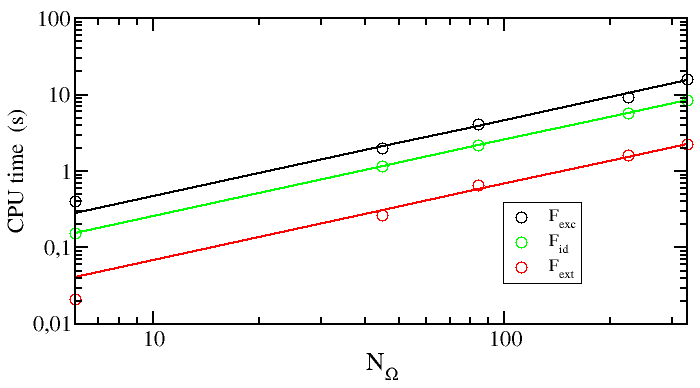}
\par\end{centering}
\caption{CPU time for the computation of the different components $F_{\textrm{id}}$,
$F_{\textrm{ext}}$, $F_{\textrm{exc}}$ of the solvation free energy
for a cubic grid of size $72^{3}$ and $n_{\textrm{max}}=3$. \label{fig:global_perf} }
\end{figure}

We show furthermore in Fig.~\ref{fig:global_perf} that despite the
complexity of the computation of $F_{exc}$ with respect to the straightforward
calculation of the local quantities $F_{ext}$ and $F_{id}$, the
computational overhead for $F_{exc}$ appears only a factor 2 with
respect to $F_{id}$ and a factor 8 with respect to $F_{ext}$. All
in all, with a sufficient grid resolution of 3 points per angstrom
and angular resolution $n_{\textrm{max}}=3$ (see below), this makes
it possible to handle, even on a single core, the solvation of small
molecules (typically $L=25$~\AA, $N\sim75$) within a minute, and
that of much larger molecules (e.g., $L=60$~\AA, $N\sim180$) in
tens of minutes. Those latter calculations were absolutely out of
reach with the direct algorithm.

\begin{figure}
\begin{centering}
\includegraphics[width=8.5cm]{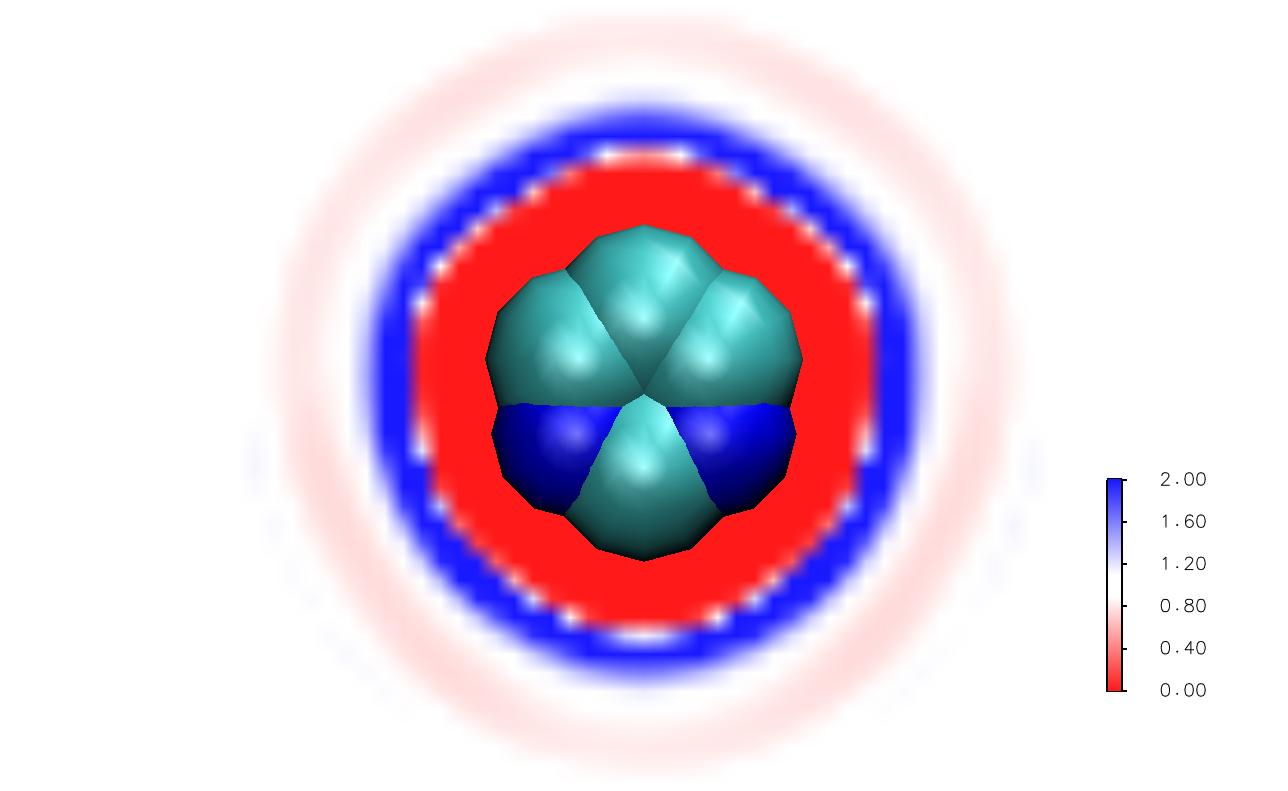}
\par\end{centering}
\begin{centering}
\includegraphics[width=8.5cm]{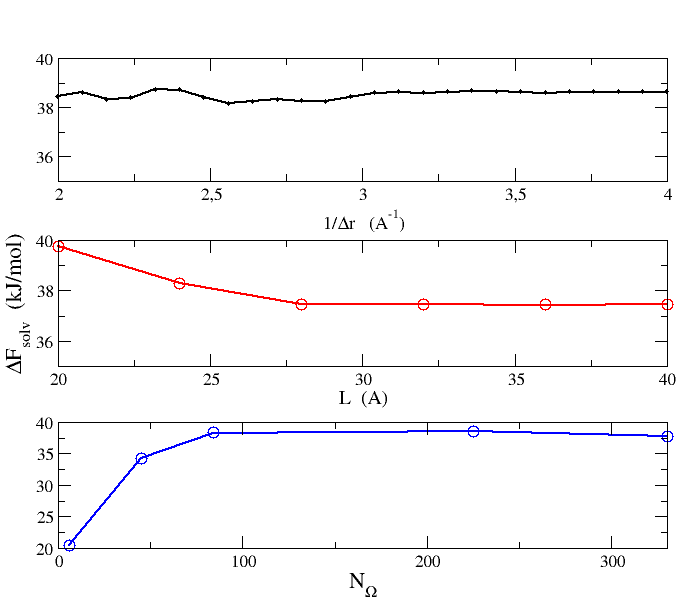}
\par\end{centering}
\caption{On top, the pyrimidine molecule with CH groups in green and N atoms
in blue. The water density map in the plane of the molecule is also
shown. Computed solvation free energy of a pyrimidine molecule in
water as a function of spatial resolution for $L=25$ \AA, $n_{\textrm{max}}=4$
(top panel), or of box size $L$ at a given resolution and number
of orientations, $\Delta r^{-1}=4$ $\textrm{�}\ensuremath{^{-1}}$,
$n_{\textrm{max}}=4$ (middle panel) , or of number of orientation
per grid point at fixed box size and spatial resolution ($L=25$ \AA,
$\Delta r^{-1}=4$ \AA$^{-1}$). In this last plot, the 5 points
are for $n_{\textrm{max}}=1$ to 5. \label{fig:sfe-pyrimidine}}
\end{figure}

In Fig.~\ref{fig:sfe-pyrimidine}, we examine the precision of the
method for the solvation free energies, taking as example a small
organic molecule, pyrimidine, dissolved in water. The three-dimensional
solvent structure resulting from the functional minimisation is shown
on top of the figure. For this neutral molecule as for many others,
we observe that in order to converge the solvation free energy, MDFT
requires a grid resolution of 3 points per Angstrom, a box length
of $28$ \AA (say a dozen of Angstrom of ``solvent buffer'' from
the molecule to the box edge in every direction), and an angular resolution
corresponding to $n_{\textrm{max}}=3$ (84 orientations). 

\begin{figure}
\begin{centering}
\includegraphics[width=8.5cm]{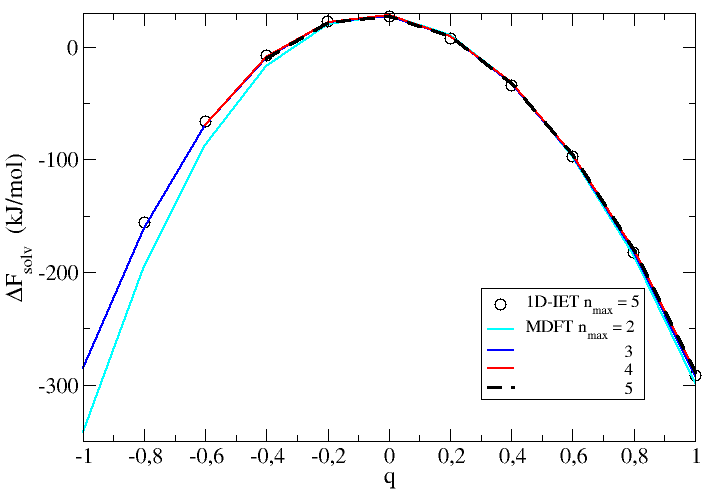}
\par\end{centering}
\caption{Solvation free energy of a hypothetical CH$_{4}^{q}$ molecule calcultated
by MDFT-HNC as a function of its charge $q$ for different angular
resolution $n_{max}=2$ to 5. The finite size corrections of equations
\ref{eq:typeB} and \ref{eq:typeC} are included. For comparison,
we also show the exact 1D-IET results that can be calculated for this
spherically symmetric case.\label{fig:SFE_CH4q}}
\end{figure}

Such results are corroborated for charged entities too, as shown in
Fig.~\ref{fig:SFE_CH4q} for the toy model corresponding to an hypothetical
CH$_{4}^{q}$ entity, that is a single Lennard-Jones center with parameters
corresponding to a unified-atoms representation of methane ($\sigma=3.73$~\AA,
$\epsilon=1.23$~kJ/mol) from Asthagiri et al. \cite{asthagiri_role_2008},
with a charge $q$ at its center. For this very specific spherically
symmetric test cases, the 1D integral equation theory is able to solve
exactly the same HNC problem, which we use as a test bed. More precisely,
MDFT results are compared to a direct integral equation resolution
of the two component system with the solute at infinite dissolution\cite{belloni14}.
This last approach implies spherical boundaries that tend toward infinity.
In the molecular density functional theory, no restriction apply to
the symmetry of the solute molecule and we use a finite box with periodic
boundary conditions. Consequently, the results of the minimisation
have to be corrected twice for charged systems\cite{kastenholz_computation_2006-1,kastenholz_computation_2006,hunenberger_single-ion_2011}.
The first correction is of the Madelung type and accounts for the
contribution of the periodic images of the solute and solvent (so-called
correction of type B)\cite{kastenholz_computation_2006}
\begin{equation}
\Delta F_{B}=-\xi(1-\frac{1}{\epsilon})\frac{q^{2}}{2L}+\mathcal{O}\left(L^{-2}\right),\label{eq:typeB}
\end{equation}
with $\xi\approx2.873$ and $\epsilon=71$ for SPC/E water \cite{berendsen_missing_1987,kusalik_science_1994}.
The other one originates from the periodic treatment of the electrostatic
potential, yielding a vanishing charge density at the box boundary
and a finite electrostatic potential in the uniform solvent (type
C)
\begin{equation}
\Delta F_{C}=-(6\epsilon_{0})^{-1}qn_{bulk}\gamma_{0},\label{eq:typeC}
\end{equation}
where $\gamma_{0}$ is the quadrupole moment of the SPC/E water molecule.

\begin{figure}
\begin{centering}
\includegraphics[width=8.5cm]{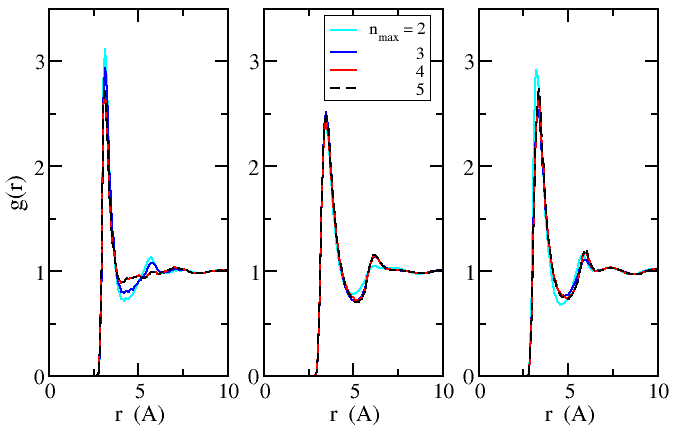}
\par\end{centering}
\caption{Reduced water density around CH$_{4}^{q}$ for different angular resolution
$n_{max}=2$ to 5 and with $q=+1$, 0 and $-0.6$ in the left, middle,
and right panels, respectively. \label{fig:gr_CH4q}}
\end{figure}

\begin{figure}
\begin{centering}
\includegraphics[width=8.5cm]{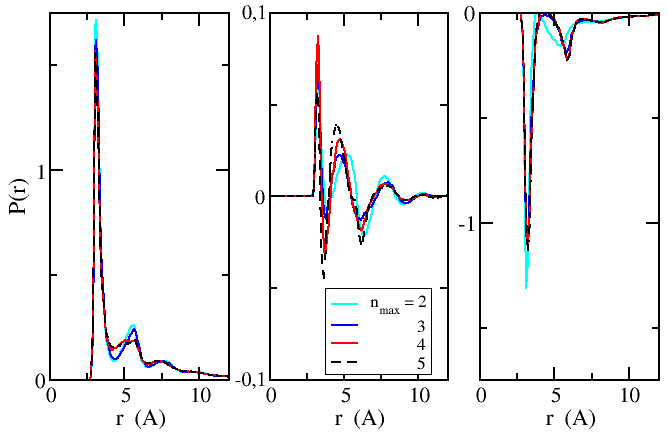}
\par\end{centering}
\caption{Polarisation density around CH$_{4}^{q}$ for different angular resolution
$n_{max}=2-5$ and with $q=+1,\:0,\:-0.6$ in the left, middle, and
right panels, respectively.\label{fig:Pr_CH4q} }
\end{figure}

In Fig.~\ref{fig:SFE_CH4q}, we show that the MDFT free energies,
including the above corrections, do match the rigorous (but spherically
symmetric only) converged HNC-IET results when the angular resolution
is increased; within the resolution of the figure, convergence of
the solvation free energy is reached for $n_{\textrm{max}}=3$. We
note that both IET and MDFT diverge for $q=-1$ and $n_{max}>3$,
a failure of the HNC approximation for this artificial solute. Fortunately,
this is not the case with Lennard-Jones parameters fitted to model
halides, e.g. from \cite{horinek_rational_2009}. In Fig.~\ref{fig:gr_CH4q}
and \ref{fig:Pr_CH4q}, we show the effect of the angular resolution
on the solvent structure for $q=+1$, $0$ and $-0.6$. We plot there
the corresponding radial distribution function (or reduced solvent
density) around the solute, $g(r)=\int d\boldsymbol{\Omega}\rho(\mathbf{r},\boldsymbol{\Omega})/n_{bulk}$
and radial solvent polarisation, $P(r)=\int d\boldsymbol{\Omega}\left(\boldsymbol{\Omega}\cdot\hat{\mathbf{r}}\right)\rho\left(\mathbf{r},\boldsymbol{\Omega}\right)/n_{\text{bulk}}$.
Although it appears that for the cationic case $q=+1$, only $n_{max}=4$
gives a full convergence of the fine structure beyond the first peak,
$n_{max}=3$ does provides overall an acceptable compromise. It is
remarquable that for the neutral case, despite a vanishing electric
field, the solute induces an expected but small finite polarisation
due to density-orientation couplings. This fine effect is slower to
converge with $n_{\textrm{max}}$. 


\begin{figure}
\begin{centering}
\includegraphics[width=8.5cm]{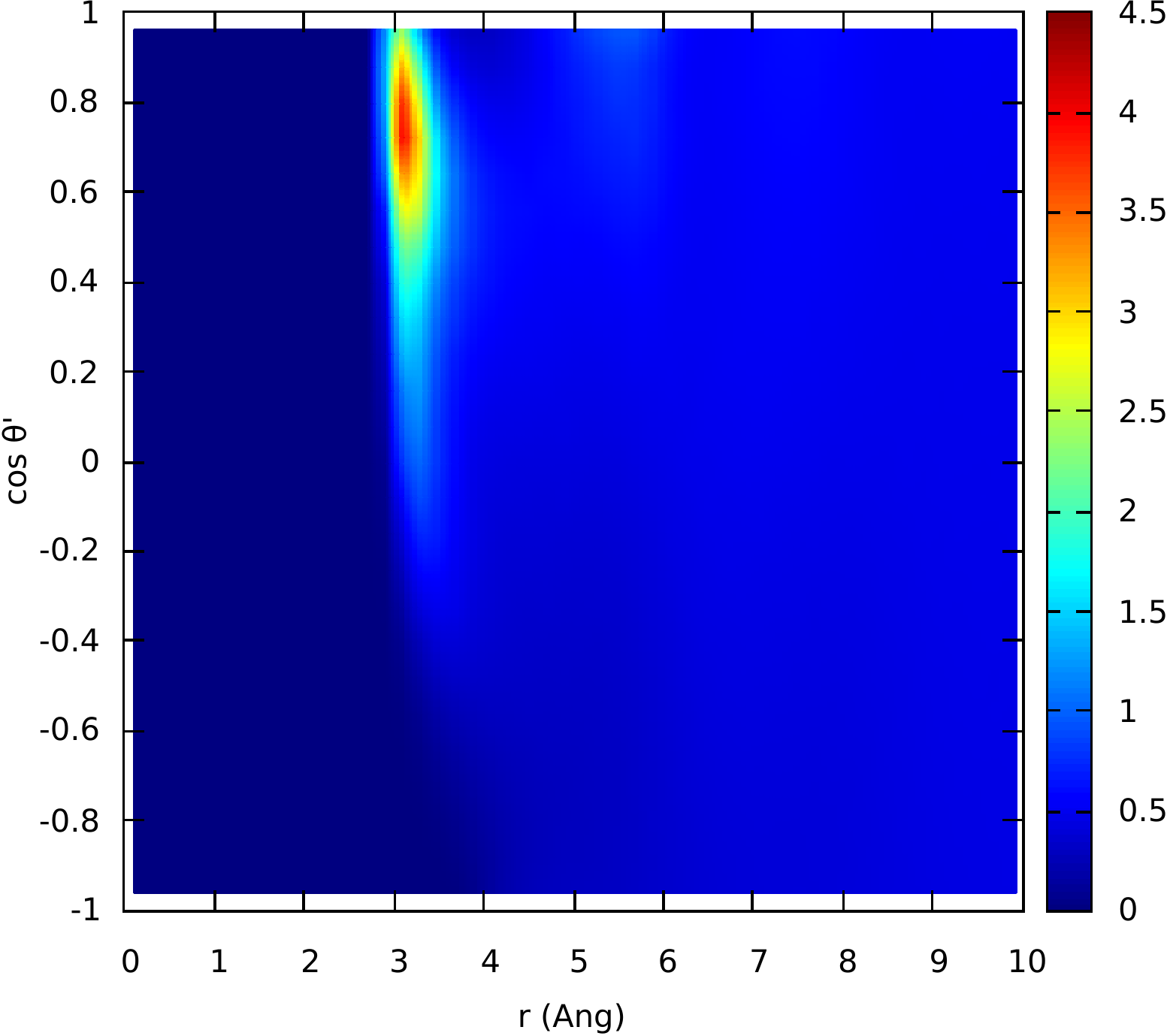}
\par\end{centering}
\caption{Distribution function between a CH$_{4}^{\text{+}}$ and a water molecule
as a function of the CH$_{4}^{+}-$O distance $r$ and the cosine
of the angle $\theta^{\prime}$ between the water dipole and the axis
joining the two sites. For each $r$ and $\cos\theta^{\prime}$, we
average over all values of intrinsic rotation angle $\psi^{\prime}$.
The distribution is not dependent of $\phi^{\prime}$ in this local
frame. This is thus a plot of $\left\langle g(r,\cos\theta^{\prime})\right\rangle _{\psi^{\prime}}$.\label{fig:g-r-costheta}}
\end{figure}

In order to illustrate the intrinsic molecular nature of the molecular
density functional theory and thus its major advantage over other
site-based liquid state theories like 3D-RISM, we show in figure \ref{fig:g-r-costheta}
the distribution function $g$ between CH$_{4}^{+}$ and the oxygen
atom of water as a function of the distance between the two sites,
$r$, and of the cosine of the angle $\theta^{\prime}$ between the
water dipole and the axis joining those sites, averaged over all intrinsic
rotations $\psi^{\prime}$. We note that in the local framework there
is invariance over angle $\phi^{\prime}$. We see that the maximum
probability is found for a distance of 3.1 \AA~and a cosine between
0.7 and 0.9. A cosine of 1 accounts for the oxygen atom pointing exactly
toward the cation. Without solvent-solvent correlations, all water
molecules would have their dipole pointing exactly toward the cation
and would thus have such cosine of 1. It is not the case here : it
is favorable to point slightly off the cation ($\cos\theta_{{\rm max}}\approx0.8$,
not 1) but to keep a more favorable short range order, i.e., to keep
more of the hydrogen bond network.

\begin{figure}
\begin{centering}
\includegraphics[width=8.5cm]{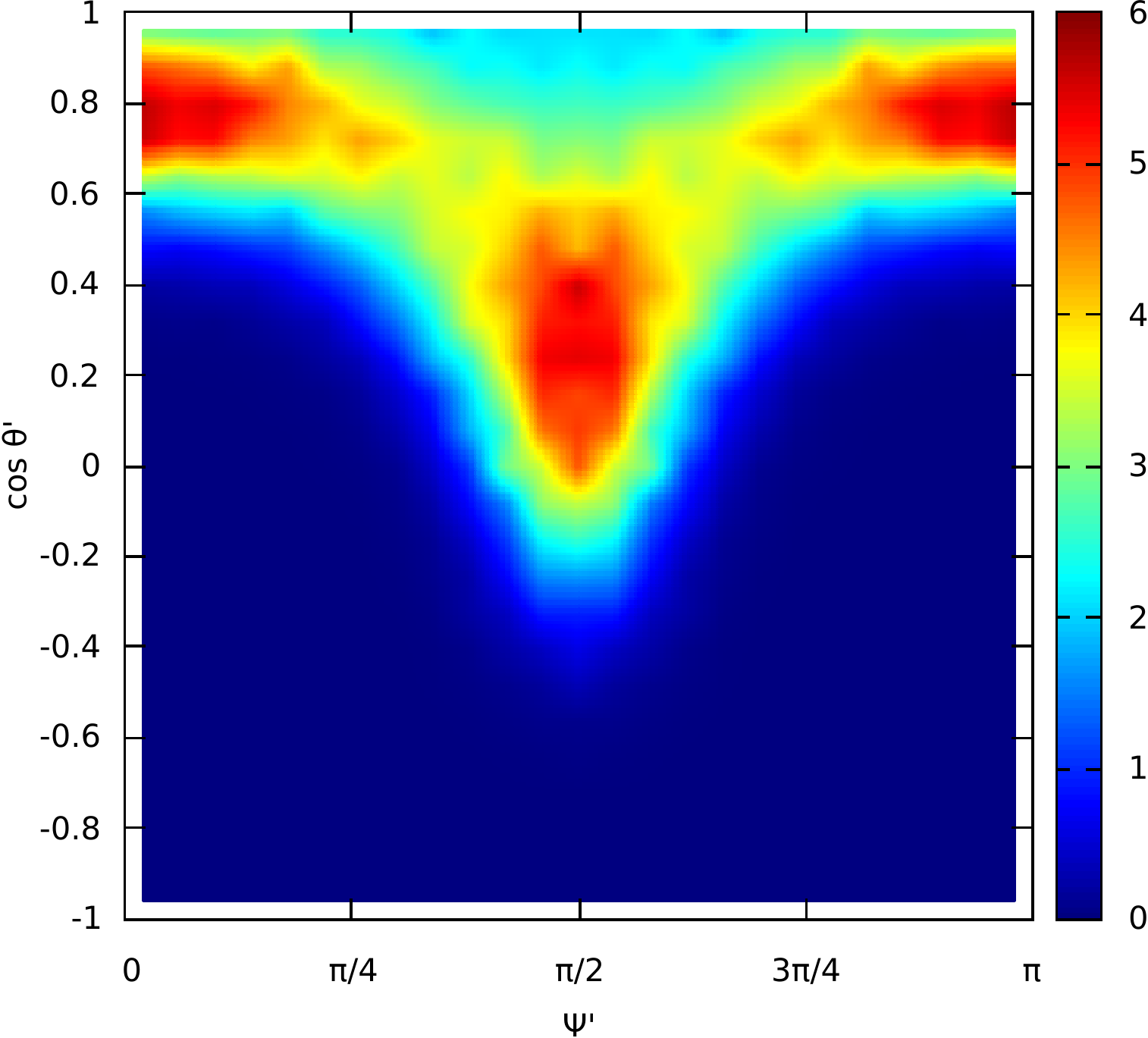}
\par\end{centering}
\caption{Distribution function between a CH$_{4}^{+}$ and the water molecule
separated by the distance $r=3.1$~\AA~as a function of the cosine
of the angle $\theta^{\prime}$ between the site-site axis and the
water dipole, and of the intrinsic rotation angle $\psi^{\prime}$.\label{fig:g-psi-costheta}}
\end{figure}

For a given distance $r=3.1$~\AA, that is in the maximum of the
radial distribution function, we show the effect of $\cos\theta^{\prime}$
and $\psi^{\prime}$ in figure \ref{fig:g-psi-costheta}. First, we
see that the distribution is symmetric around $\psi^{\prime}=\pi/2$,
as expected from a C$_{2{\rm v}}$ molecule in the reference framework.
The maximum of the distribution is again found for $\cos\theta^{\prime}$
between 0.65 and 0.75, that is for a dipole pointing roughly toward
the cation. For a dipole perpendicular to the solute-oxygen vector,
that is for $\cos\theta^{\prime}=0$, we see that the internal rotation
of $\psi^{\prime}=\pi/2$ that produces the two hydrogen the farthest
from the cation is much more probable that other internal rotations.

\begin{figure}
\begin{centering}
\includegraphics[width=8.5cm]{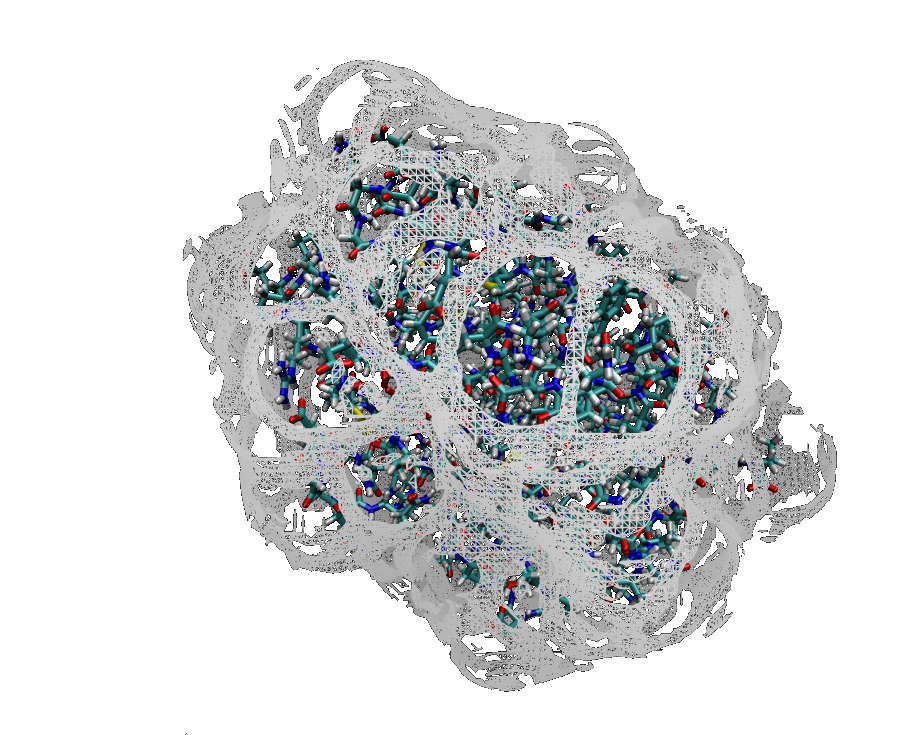}
\par\end{centering}
\caption{Water density around a protein made of 230 residues and 4000 atomic
sites (4M7G: Streptomyces Erythraeus Trypsin). The displayed isosurface
correspond to 3 times the bulk density. \label{fig:4M7G}}
\end{figure}

We conclude by a proof of concept to show that this formalism is efficient
enough to unlock the description of the solvation around large molecular
solutes. In Fig.~\ref{fig:4M7G}, we show the water structure around
a protein made of 4000 atoms corresponding to 230 residues. We use
a grid of $128^{3}$ nodes, an angular resolution corresponding to
$n_{\textrm{max}}=3$ and a discretization of the grid of $0.5$~\AA.
The overall minimisation took 2 minutes on 24 distributed cores. Using
MD simulations, an equivalent statistics for the water density requires
at least 100 ns and hundreds of cpu-hours with the same computer ressources.
It would be even more challenging to get the whole angular-dependent
density $\rho(\mathbf{r},\boldsymbol{\Omega})$, a direct output of
the functional approach. 

\section{Conclusion}

The three-dimensional density functional theory and integral equation
formalism at the molecular level of description of the solvent has
been greatly improved by using the concept of expansions/projections
onto generalized spherical harmonics. The present analysis of the
Ornstein-Zernike convolution product follows that previously developed
in bulk systems. The resulting algorithm decreases the time-to-solution
by many orders of magnitude. This makes it possible to study in a
systematic and routine way many solute/solvent mixtures and to provide
free energies of solvation with restitution times of at most a few
minutes. Applications to simple molecular solutes in water have been
presented. A detailed assessment of the method with respect to reference
MD calculations or experimental data, as well as examination of large
molecular systems of biological interest, like the prediction of protein
hydration, will be reported soon in a companion paper.

The general algorithm presented in this paper could be accelerated
following different directions \textendash not speaking of making
it highly parallel. First, it is important to note that the $\gamma$
function is a convolution product. It is thus smoother (both in spatial
and angular dependence) than its two building blocks $\Delta\rho$
and $c$. As a consequence, it is legitimate to use a degraded basis
$\left\{ n_{\max}'\right\} $ with $n_{\max}'<n_{\max}$ during the
entire process. Also, inhomogeneous grids in space and orientations
seem logical extensions to the important milestone reported therein.
These technics may lead to further substancial decrease in time-to-solution
without altering the precision.

Now that the numerical barrier has been unlocked remains an important
question. As usual in such liquid-state theories, the validity of
the HNC-like DFT functional has to be challenged, and one will have
to go beyond this approximation by building solute-solvent bridge
function(al)s. We already have several suggestions in that directions,
either based on global thermodynamic corrections \cite{levesque12_1,jeanmairet_molecular_2013,jeanmairet_molecular_2013-1,gageat_coarseGrainedBridge_2017}
or on a detailed understanding of the bridge functions for simple
molecular systems\cite{belloni12}. 

\section*{Appendix: angular representation versus projections}

The expansion (\ref{eq:1.9}) and the projection (\ref{eq:1.11})
which transform triplets of angles $\mathbf{\Omega}\equiv(\theta,\phi,\psi)$
to indices $_{\mu'\mu}^{m}$ or vice versa follow a three-step algorithm
originally developed for bulk systems\cite{lado_95}:

First and second steps: transform $\phi$ and $\psi$ into $\mu'$
and $\mu$:
\begin{equation}
\Delta\rho_{\mu'\mu}(\theta)=\dfrac{1}{4\pi^{2}}\int_{0}^{2\pi}\int_{0}^{2\pi}\Delta\rho(\theta,\phi,\psi)e^{+i\mu'\phi+i\mu\psi}\mathrm{d}\phi\mathrm{d}\psi\label{eq:1.33}
\end{equation}
\begin{equation}
\Delta\rho(\theta,\phi,\psi)=\sum_{\mu'=-n_{\max}}^{n_{\max}}\sum_{\mu=-n_{\max}}^{n_{\max}}\Delta\rho_{\mu'\mu}(\theta)e^{-i\mu'\phi-i\mu\psi}\label{eq:1.33-2}
\end{equation}
The 2D angular integral \ref{eq:1.33} is performed by trapezoidal
rule (or Gauss Chebychev quadrature):
\begin{equation}
\Delta\rho_{\mu'\mu}(\theta)=\dfrac{1}{N_{\phi}N_{\psi}}\sum_{j=0}^{N_{\phi}-1}\sum_{k=0}^{N_{\psi}-1}\Delta\rho(\theta,\phi_{j}\equiv j\dfrac{2\pi}{N_{\phi}},\psi\equiv k\dfrac{2\pi}{N_{\psi}})e^{+2i\pi\left(\frac{\mu'j}{N_{\phi}}+\frac{\mu k}{N_{\psi}}\right)}
\end{equation}
One recognizes a discrete 2D Fourier transform which can be efficiently
performed by 2D FFT, provided $N_{\phi}=N_{\psi}=2n_{\max}+1$. Same
remark for the inverse transformation \ref{eq:1.33-2}. The case of
${\rm H_{2}O}$ symmetry can be adapted by choosing $N_{\psi}=2\left(n_{\max}/2\right)+1$
angles between 0 and $\pi$ . This operation must be performed for
each $\theta$ value. 

Third step: transformation between $\theta$ and $m$:
\begin{equation}
\Delta\rho_{\mu'\mu}^{m}=f_{m}\int_{-1}^{1}\dfrac{\mathrm{d}\cos\theta}{2}\Delta\rho_{\mu'\mu}(\theta)r_{\mu'\mu}^{m}(\theta)=f_{m}\sum_{i=1}^{N_{\theta}}w_{i}\Delta\rho_{\mu'\mu}(\theta_{i})r_{\mu'\mu}^{m}(\theta_{i})
\end{equation}
\begin{equation}
\Delta\rho_{\mu'\mu}^{m}=\sum_{m=\max(\left|\mu'\right|,\left|\mu\right|)}^{n_{\max}}\Delta\rho_{\mu'\mu}^{m}r_{\mu'\mu}^{m}(\theta)
\end{equation}
The integral over $\theta$ is performed using Gauss-Legendre quadrature
with $N_{\theta}=n_{\max}+1$ and associated weights $w_{i}$. It
is performed for each pair $\left\{ \mu',\mu\right\} $.

Despite the lack of \textquotedbl{}Fast\textquotedbl{} transform in
this last step, the whole procedure is fast enough not to be the limiting
process in the OZ convolution calculation. Overall, we can qualify
the whole angles-to-projections process, analogous to a FFT for the
angular variable, as a Fast Generalized Spherical Harmonics Transform
(FGSHT).
\begin{acknowledgments}
This work was supported by the Energy oriented Centre of Excellence
(EoCoE), grant agreement number 676629, funded within the Horizon2020
framework of the European Union.
\end{acknowledgments}

\bibliographystyle{unsrt}
\bibliography{main}

\end{document}